%% file: aaai23.tex
\documentclass{article}
\usepackage{arixv}
\usepackage{multirow}
\usepackage{graphicx}
\usepackage[utf8]{inputenc} 
\usepackage[T1]{fontenc}    
\usepackage{babel}
\usepackage{hyperref}       
\usepackage{url}            
\usepackage{booktabs}       
\usepackage{amsfonts}       
\usepackage{nicefrac}       
\usepackage{microtype}      
\usepackage{lipsum}		
\usepackage{natbib}
\usepackage{doi}
\usepackage{amsmath}
\usepackage{bm}
\usepackage{booktabs} 
\usepackage{float}
\usepackage{amsmath}
\usepackage{amssymb}
\usepackage{authblk}
\usepackage{appendix}
\title{GOLLIC: Learning Global Context beyond Patches for Lossless High-Resolution Image Compression }

\author{
    {\normalsize\textbf{Yuan ~Lan}}\textsuperscript{\rm 1,\rm 2},
    {\normalsize\textbf{Liang ~Qin}}\textsuperscript{\rm 1},
    {\normalsize\textbf{Zhaoyi ~Sun}}\textsuperscript{\rm 1},
    {\normalsize\textbf{Yang ~Xiang}}\textsuperscript{\rm 2,\rm 3}\thanks{Corresponding author},
    {\normalsize\textbf{Jie ~Sun}}\textsuperscript{\rm 1}\thanks{Corresponding author}}
\affil{
    \textsuperscript{\rm 1}Theory Lab, Huawei Hong Kong Research Center\\
\textsuperscript{\rm 2}Department of Mathematics, The Hong Kong University of Science and Technology\\
\textsuperscript{\rm 3}Algorithms of Machine Learning and Autonomous Driving Research Lab, HKUST Shenzhen-Hong Kong Collaborative Innovation Research Institute.\\
ylanaa@connect.ust.hk,
\{qin.liang, sun.zhaoyi1, j.sun\}@huawei.com,
maxiang@ust.hk
}

\hypersetup{
pdftitle={GOLLIC-2022},
pdfsubject={q-bio.NC, q-bio.QM},
pdfauthor={Yuan Lan},
pdfkeywords={Lossless Compression, Self-supervised Clustering, Deep Learning},
}

\begin{document}
\maketitle

\begin{abstract}

\input{./content/abstract.tex}
\end{abstract}

\input{./content/introduction.tex}

\input{./content/related_work.tex}
\input{./content/background.tex}

\input{./content/method.tex}

\input{./content/experiment.tex}

\input{./content/conclusion.tex}
\input{./content/acknowledge.tex}
\input{./content/appendix.tex}

\bibliographystyle{unsrtnat}
\bibliography{aaai23}

\end{document}

%% file: content/abstract.tex
Neural-network-based approaches recently emerged in the field of data compression and have already led to significant progress in image compression, especially in achieving a higher compression ratio. In the lossless image compression scenario, however, existing methods often struggle to learn a probability model of full-size high-resolution images due to the limitation of the computation source. The current strategy is to crop high-resolution images into multiple non-overlapping patches and process them independently. This strategy ignores long-term dependencies beyond patches, thus limiting modeling performance. To address this problem, we propose a hierarchical latent variable model with a global context to capture the long-term dependencies of high-resolution images. Besides the latent variable unique to each patch, we introduce shared latent variables between patches to construct the global context. The shared latent variables are extracted by a self-supervised clustering module inside the model's encoder. This clustering module assigns each patch the confidence that it belongs to any cluster. Later, shared latent variables are learned according to latent variables of patches and their confidence, which reflects the similarity of patches in the same cluster and benefits the global context modeling. Experimental results show that our global context model improves compression ratio compared to the engineered codecs and deep learning models on three benchmark high-resolution image datasets, DIV2K, CLIC.pro, and CLIC.mobile.

%% file: content/introduction.tex
\section{Introduction}
In recent years, there has been an explosive growth in the number of high-resolution digital images due to the rapid development of multimedia technology. 
The increasing volume of images poses a challenge to the system's transmission bandwidth and storage capacity. Image compression algorithms \cite{Pennebaker1992JPEGSI,Boutell1997PNGN,sayood2017introduction} address such challenges by compressing images into smaller-sized bit streams. Although lossy compression can bring a much higher compression rate, lossless image compression, which allows the image to be reconstructed perfectly without any loss, is still highly demanded in the business scenario, such as storage service providers. Hence, developing a higher-performance lossless image compression algorithm is necessary to cope with the growing demand for high-resolution image storage.
\begin{figure}
    \centering
\includegraphics[width=0.5\textwidth]{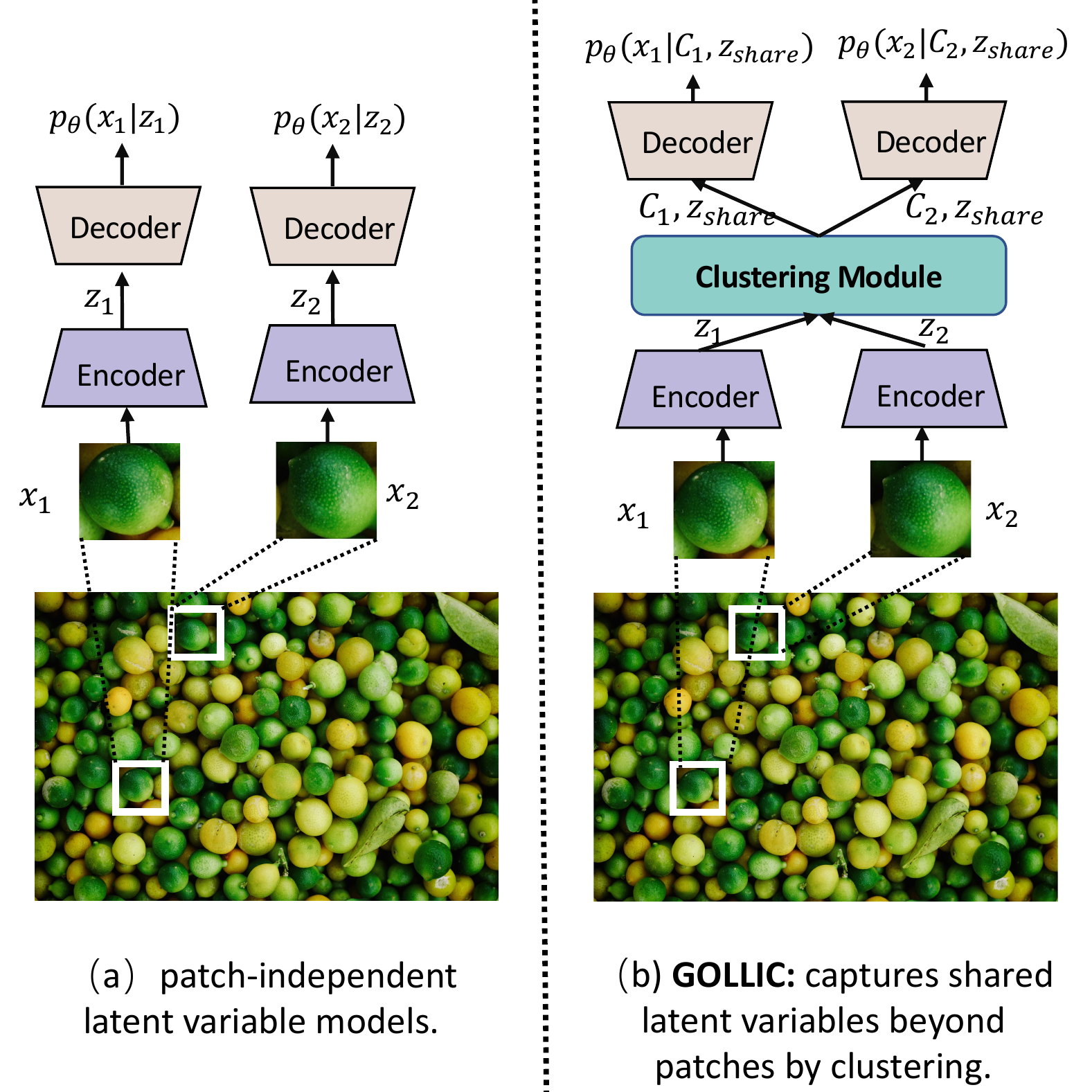}
    \caption{An example to illustrate the difference between GOLLIC and patch-independent models. To model two similar patches $x_1$, $x_2$, (a) the patch-independent models process them conditioned on independent latent variables $z_1$, $z_2$. (b) our proposed GOLLIC captures their dependencies by a clustering module, then models them with their shared latent variable $z_{share}$ and their confidence in each cluster $C$.}
    \label{fig:gollic_summary}
\end{figure}


Typical lossless image compression models consist of a probability model and an entropy coder. The probability model describes the image's statistical characteristics and lays the foundation of the following encoding process. In the encoding process, the entropy coder maps the image to a sequence of code words with a smaller length using information from the probability model. In Shannon's source coding theory \cite{shannon1948mathematical}, the probability distribution of data determines the theoretical limit of compression rate. Similarly, an accurate probability model is crucial to achieving a high compression rate.

There are two branches of lossless compression algorithms: the traditional engineered algorithms and deep learning models. Among the traditional engineered lossless compression methods, predictive coding \cite{sayood2017introduction} is one popular scheme adopted in formats such as JPEG2K \cite{rabbani2002book}, JPEG-LS \cite{weinberger2000loco}. 
The main idea of predictive coding is to reduce the spatial correlation between pixels by a predictor hence simplifying the probability modeling process. One type of common-used predictor is the context-based adaptive predictor such as the edge-directed prediction (EDP) \cite{li2001edge} and the gradient-adaptive predictor (GAP) used in CALIC \cite{wu1997context}. 
Those predictors utilize information about each pixel's local neighborhood, such as gradient and the direction of the edge, to model the spatial correlation between pixels. 
Although those context-adaptive schemes are easy to implement, they ignore the inherent long-term dependencies in natural images
 \cite{Dabov2007ImageDB} between pixels not in close neighborhoods. There are non-local predictors addressing this problem \cite{wu2010adaptive,chen2013nonlocal,jing2013lossless,crandall2014bm3d, novikov2016local} by searching long-range patterns in the image. However, such pattern searching is computationally heavy, especially for high-resolution images.

Deep neural network methods show their power in lossless compression recently \cite{van2016conditional,townsend2019hilloc,mentzer2019practical,cao2020lossless,hoogeboom2019integer}. Unlike engineered compression codecs that manually design the probability model based on statistical information on image, deep neural network based methods model the probability distribution of image directly. Thanks to the powerful capacity for approximating  complicated distribution, deep neural network-based models generally achieve higher compression rates than the traditional compression codecs. One of the state-of-the-art models are autoregressive model \cite{van2016conditional,van2016pixel,salimans2017pixelcnn++}. They model the probability of image pixels in sequential order as Markov chains. However, its sequential nature slows down the encode and decode speed, making autoregressive models impractical for processing high-resolution images. Another type of deep learning model is hierarchical latent variable models \cite{mentzer2019practical,cao2020lossless}. They parallelly derive the probability distribution of all image pixels by conditioning on latent variables. It speeds up the encoding and decoding process and makes the neural network-based compressor practical for high-resolution images. However, in practice, one must crop high-resolution images into multiple fixed-size patches due to the limitation of GPU memory and model architecture. Therefore the long-term dependencies among patches are ignored.

To capture the long-term dependencies beyond patches, we propose a hierarchical latent variable model with the global context. We name this  \textit{\textbf{G}lobal c\textbf{O}ntext \textbf{L}earned \textbf{L}ossless \textbf{I}mage \textbf{C}ompression model} as GOLLIC.
Unlike previous methods that treat each patch's latent variables as independent, we introduce shared latent variables among patches to capture similar contents. Fig.\ref{fig:gollic_summary} give an example to illustrate this idea. Then those shared latent variables construct a global context for probability modeling. To obtain the shared latent variables, we equip the model's encoder with a self-supervised clustering module. During training, this clustering module assigns each patch the confidence that it belongs to any cluster. Later, those learned confidence weights each patch's latent variable to generate the shared latent variables. Usually, patch searching methods in non-local predictive coding cluster similar patches to capture long-term dependencies by explicitly defining a distance and comparing patches' distances, which brings intensive computational complexity. Instead, our clustering module does not define the distance of patches. It directly estimates the probability of patches to each cluster by the encoder, thus avoiding the computational complexity brought by patch searching. Also, we set an auxiliary objective function that guides the shared features to reconstruct the last level's latent variables. This auxiliary objective function allows the clustering module to avoid cluster collapse and demonstrate stable performance during training.
We summarize the highlights of GOLLIC as follows,
\begin{itemize}
    \item GOLLIC considers the shared latent variables among patches and uses them to construct a global context, thus benefiting the accurate probability modeling.
    \item The self-supervised clustering module introduced to obtain the shared latent variables is easy to use. It does not bring extra computational complexity like the traditional non-local patch searching method.
    \item Our experimental result shows that our method improves compression ratio compared to multiple engineered codecs and latent variable models on three high-resolution image datasets.
\end{itemize}

%% file: content/related_work.tex
\section{Related Works}
\subsection{Hierarchical Latent Lossless Compression}

Latent variable models are one of the mainstream neural-network-based lossless compression models. They assume that data are generated by latent variables and aim to learn such latent variables. Given such learned latent variables, one could model data's probability distribution more accurately, achieving a better compression rate. In recent literature, \cite{kingma2019bit,mentzer2019practical,cao2020lossless}, models using hierarchically represented latent variables show success in improving the compression rate. L3C \cite{mentzer2019practical} utilizes three-level hierarchical latent variables, in which each level's latent variables model the previous level's latent variables. Similarly, SReC \cite{cao2020lossless} designs a three-level hierarchical latent variables model inspired by super-resolution techniques. In addition, Bit-Swap \cite{kingma2019bit} combines bits-back coding with hierarchical latent variables by Markov chains to achieve better modeling performance on high-dimensional distributions. However, in practice, high-resolution images are divided into fixed-size patches. These previous works ignore the dependency among patches' latent variables. Instead, our work utilizes hierarchical latent variables with a clustering structure that captures the latent variables' dependency shared among patches.

\subsection{Long-term Dependencies Modeling}
Over the past decades, many works have discussed the long-term dependencies of data for compression. Usually, long-term dependencies inside the image indicate similar patterns but are not in the near neighborhood. One typical method called non-local means (NLM) \cite{buades2005non, Dabov2007ImageDB} first succeeded in image denoising, then extended the idea to lossless image compression and improved the performance of predictive coding. Those methods emphasize the self-similarity structures inside images rather than the local spatial correlation of images. For example, In \citeauthor{chen2013nonlocal}'s work, they capture long-term dependencies using the statistical information of similar patches to calculate a context number for each pixel. The pixels with the same context number were grouped into one class and processed together. Besides this, the character of similar patches was also used to predict the value of pixels directly \cite{crandall2014bm3d,novikov2016local}. However,  searching similar patches in all pixels leads to extra computational complexity, which hinders such algorithms from being widely used in engineered formats.

Recently, some deep learning models have utilized the long-term dependencies to improve learned lossy compression \cite{chen2019neural,li2021learning,liu2019non,patel2021saliency}. Nevertheless, in deep neural network-based lossless compression, the non-local dependencies have not been widely explored yet.



%% file: content/background.tex
\section{Background}
\subsection{Lossless Compression}
In this section, we review the basics of lossless compression.
The goal of the lossless compression task is to compress the data from the source into a sequence of code words with smaller code lengths in a reversible manner. 
Specifically, given source data $\bm{X} = \left\{x_1,\cdots,x_N\right\},  x \in \mathcal{A}$, where $\mathcal{A}$ is a finite alphabet, $\bm{X}$ follows $p(\bm{X})$, the entropy coder aims to find a mapping from source alphabet $\mathcal{A}$ to a finite code alphabet $\mathcal{C}$ such that the source data can be recovered exactly from the code words in $\mathcal{C}$.
Meanwhile, an optimal entropy coder is also required to minimize the expected length of the code words.As stated in Shannon's source coding theorem \cite{shannon1948mathematical}, the optimal expected code length is equal to entropy $H(\bm{X}) = - p(\bm{X})log_2 p(\bm{X})$.
However, in practice, $p(\bm{X})$ is usually intractable. Hence, the target of probability modeling is to approximate $p(\bm{X})$ as close as possible. One way to to factorize $p(\bm{X})$ is using the independence assumption. The simplest model is to assume each $x_i$ in $\bm{X})$ is statistically independent, then $p(\bm{X}) \approx \prod_i^{N}p(x_i)$.  
Besides, the more accurate way to approximate  $p(\bm{X})$ is context-based model, i.e.
\begin{equation}\label{eq:context-model}
    p(\bm{X}) \approx \prod_i^{N}p(x_i|B_i)
\end{equation}
where context $B_i = \left(x_{i-1},\cdots,x_{i-m}\right), m\leq i$.
When context size $m$ is larger, we obtain more accurate modeling and more computational burdens. Therefore, extending the context efficiently is a challenge in the probability model of lossless compression.

\subsection{Latent Variable Model for Lossless Compression}

Context is essential to accurately approximate the actual probability distribution of data $p(\bm{X})$, such as $B_i$ consists of symbols in the source data. Nevertheless, one could construct contexts beyond symbols in the source data. One of such general contexts is the latent variable. In particular, latent variables $\bm{z}$ are statistical summarizations of the data $X$ that allow us to approximate the conditional probability $p(X|\bm{z})$ similar to Eq. \ref{eq:context-model}
\begin{equation}\label{eq:latent}
    p(\bm{X}|\bm{z}) \approx \prod_i^{N}p(x_i|\bm{z})
\end{equation}
The major difference between latent variable $\bm{z}$ and context $B_i$ in Eq. \ref{eq:latent} is that $\bm{z}$ are not necessarily symbols from the source data. The price of such generalization is that, apart from learning the probability distribution $p(\bm{X}|\bm{z})$, one has to train an encoder $q(\bm{z}|\bm{x})$ that summarizes the data $X$ into latent variables $\bm{z}$. In practice, both $p(\bm{X}|\bm{z})$ and $q(\bm{z}|\bm{x})$ are parametrized by deep neural networks with parameters $\theta$,$\Phi$ which are estimated by maximizing the evidence lower bound objective (ELBO).
The benefit of latent variables is that both $p(\bm{X}|\bm{z})$ and $q(\bm{z}|\bm{x})$ could be evaluated parallelly, while the context $B_i$ in the previous section must be evaluated in a sequential order.







\subsection{Predictive Coding} 
The predictive coding scheme is one of the most efficient techniques in engineered lossless compression codecs \cite{sayood2017introduction}. 
It adds a prediction step before building the probability model, rather than deep learning-based approaches that model the probability distribution directly. The prediction step is to predict the value of the current pixel by the processed neighboring pixels, which efficiently decorrelates the spatial correlation between neighboring pixels. Hence, the prediction errors can be modeled easily. 

Our work employs the Median Edge Detector (MED) to decorrelate the local spatial dependency. MED detects the edge based on local context, and its prediction for pixel $x_{ij}$ can be written as follows, where 
\begin{equation}
\label{eq:MED}
\hat{x}_{ij}=\left\{
\begin{aligned}
\text{min}(a,b) & , & \text{if} \, c\geq \text{max}(a,b), \\
\text{max}(a,b) & , & \text{if}\, c \leq \text{min}(a,b) \\
a+b-c & , &\text{otherwise}.
\end{aligned}
\right.
\end{equation}
where $i,j$ indicate the location of row and column, respectively. $a,b,c$ denote the pixels $x_{i,j-1},x_{i-1,j}$ and $x_{i-1,j-1}$, respectively.
Compared to linear prediction, MED is much more flexible in adapting to smooth and edge regions of images. However, the limited size of support pixels set $a,b,c$ prevents them from improving the performance further. Therefore, the following model which captures the longer dependencies is necessary.


%% file: content/method.tex
\section{Method}
\begin{figure*}[h]
    \centering
    \includegraphics[width=1\textwidth]{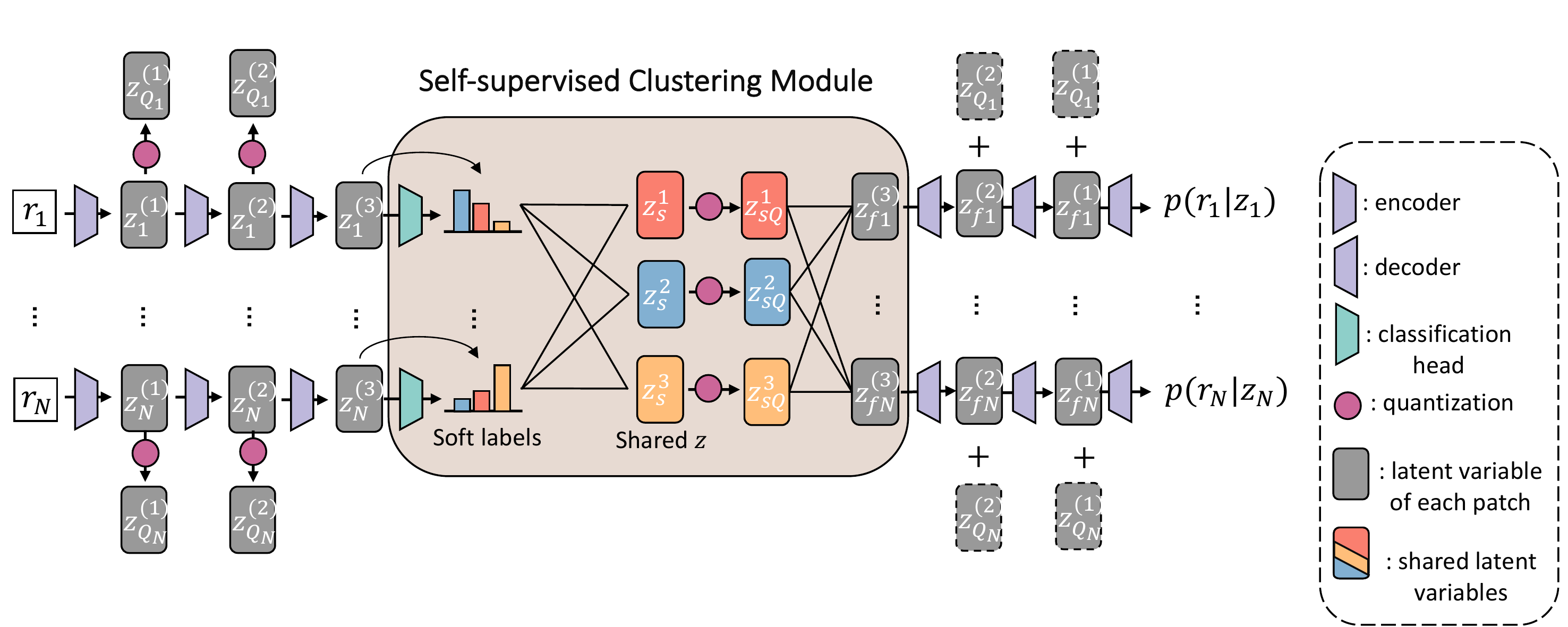}
    \caption{Overview of our GOLLIC's Framework. The hierarchical framework introduces latent variables on three scales, accompanied by three layers of encoders and decoders. In addition, we embed extra clustering modules into the encoders to obtain the shared latent variables and the probabilities of patches in each cluster (denoted as the soft label in the figure). For simplicity, inside $p(\bm{r_i}|\bm{z_i})$, we use $\bm{z_i}$ to denote all latent variables used in decoding. }
    \label{fig:gollic_framework}
\end{figure*}

In this section, we introduce our proposed method, GOLLIC. It is a hierarchical latent variable model used for lossless image compression. To capture the global context beyond patches, we embed a self-supervised clustering module into the encoder. We describe this clustering module in the encoder and how GOLLIC works as follows.

\subsection{Preprocessing}
We first apply reversible color transforms and a predictive coding predictor to the original high-resolution image $\bm{X}\in [0,255]^{3 \times H \times W}$, reducing the local correlations among pixels. The reversible color transforms convert RGB channels to YCrCb by the reversible integer transform \cite{gormish1997lossless} as in the JPEG2000 format. This color transform aims to decorrelate the correlation between RGB channels efficiently. Then we apply the Median Edge Detector Eq. \eqref{eq:MED} as the predictive coding predictor to decorrelate pixels. The predictor subtracts from each pixel its predicted value according to the pixel's neighborhoods. Therefore the processed picture consists of the prediction residuals having the same size as $\bm{X}$ with values in the range of $[-255,255]$. Then we mod the processed picture by 256 to rescale its values in the $[0,255]$ range. 
Finally, the processed picture is cropped into multiple non-overlap patches with size $N \times N$. Those patches are stacked into a 4-dim tensor in raster-scan order. We denote  this 4-dim  tensor as $\bm{R} = \left(\bm{r}_1, \cdots,\bm{r}_P\right)^T \in [0,255]^{P \times 3\times N \times N}$, where $\bm{r} \in [0,255]^{3\times N \times N}$ denotes patch, $P$ is the number of patches. Next, we introduce our proposed model approximating the probability distribution of $\bm{R}$.

\subsection{Architecture }
Fig.\ref{fig:gollic_framework} depicts the framework of our GOLLIC model. The GOLLIC model utilizes a three-level hierarchical framework based on the L3C \cite{mentzer2019practical} latent variable model. The hierarchical framework introduces latent variables on three different scales, accompanied by three layers of encoders and decoders. In addition, we embed extra clustering modules into the encoders to obtain the shared latent variables. Therefore our model consists of three parts: encoders, clustering modules, and decoders.
In practice,  input of the encoder is processed picture consists of stacking patches prediction residuals from predictive coding prediction $\bm{R}= \left(\bm{r}_1, \cdots,\bm{r}_M\right)^T$. The three layers of encoders extract from $\bm{R}$ the hierarchical latent variables $\bm{z}^{(n)}$ with size $P\times C_f\times \frac{N}{2^n}\times \frac{N}{2^n}$, $n=1,2,3$ respectively. Then the final layer of hierarchical latent variable $\bm{z}^{(3)}$ is fed into the self-supervised clustering module for further clustering of patches.
In the clustering module, the classification head group patches into clusters according to the latent variable $\bm{z}^{(3)}$. It outputs each patch's probability of belonging to different clusters. These  probabilities are then used to weight $\bm{z}^{(3)}$ to obtain the shared latent variables $\left\{\bm{z_s}^{i}\right\}, i=1,\cdots,K$ (see the details in the next section).
To store the shared latent variables $\bm{z}^{(1)}$,  $\bm{z}^{(2)}$ and $\left\{\bm{z_s}^{i}\right\}$, we quantize them by the scalar differentiable quantization function in  \cite{mentzer2019practical}. 
In the decoding process, we reconstruct the latent variable $\bm{z_f}^{(3)}$ of each patch through summation of the quantized shared latent variables $\left\{\bm{z_{sQ}}^{i}\right\}$ weighed by their soft labels. The decoder then upsample $\bm{z_{f}}^{(3)}$ to $\bm{z_{f}}^{(2)}$, which is added with the quantized latent variable $\bm{z}_Q^{(2)}$ to reconstruct $\bm{z_f}^{(1)}$. Similarly, $\bm{z_f}^{(1)}$ are added with the quantized latent variable $\bm{z}_Q^{(1)}$ in the first-layer to estimate $p(\bm{r}|\bm{z})$ after upsampling.

\subsection{Self-Supervised Clustering Module}
In this section, we discuss the self-supervised clustering module in our architecture in detail. Our self-supervised clustering module aims to describe the global context by extracting shared features between latent variables of image patches. The self-supervised clustering module is equipped at the bottom of the hierarchical network. It splits the latent variable $\bm{z}^{(3)}$ into two parts: the soft labels and the shared latent variables $\bm{z_{s}}^i$ with $i=1,\cdots, K$. The soft labels describes each patch’s probability of belonging to $K$ clusters. The shared latent variables are the global context targeted to summarize the common characteristics of patches belonging to each cluster.

Firstly, we discuss how our self-supervised clustering module generates the soft labels for each patch. The soft labels is generated by latent variables at the third level of the hierarchical framework. We denote these latent variables as $\bm{z}^{(3)}\in \mathbb{R}^{P \times C_f\times \frac{N}{2^3} \times \frac{N}{2^3}}$, where $P$ are the number of patches and $C_f$ denotes the channel of the latent variable. To obtain the probabilities of patches to $K$ clusters, $\bm{C} \in [0,1]^{P\times K}$, we pass $\bm{z}^{(3)}$ through a classification head. This classification head is a nonlinear transform and consists of three components: a layer $f_d = Conv^{(2)}\sigma\circ Conv^{(1)}$ which map $\bm{z}^{(3)}$ into the feature with size $P \times C_d\times \frac{N}{2^3} \times \frac{N}{2^3}$, a fully-connected layer $f_m$ that maps features from $P \times C_d\times \frac{N}{2^3} \times \frac{N}{2^3}$ into $P\times K$,  and a softmax layer. Hence, the clustering probabilities $\bm{C}$ are formulated as follows,
\begin{equation}\label{eq:C}
    \bm{C} = Softmax(f_m(f_d(\bm{z}^{(3)})))
\end{equation}

Secondly, we discuss how our self-supervised clustering module generates shared latent variables. After obtaining the clustering probabilities $\bm{C}$, we obtain the shared latent variables $\bm{z_{s}}$ by weighted grouping $\bm{z}^{(3)}$ according to $\bm{C}$. Explicitly, we map $\bm{z}^{(3)}$ into the feature $\bm{h}^{3}\in \mathbb{R}^{P \times C_d\times \frac{N}{2^3} \times \frac{N}{2^3}}$ by a ResNet block. The feature $\bm{h}^{3}$ is reshaped into 2-dim matrix with size $P\times L$, where $L$ denotes $C_d\times \frac{N}{2^3} \times \frac{N}{2^3}$. Then we obtain the shared latent variables $\bm{z_{s}}$ by weighted summation of $\bm{h}^{3}$ w.r.t clustering probabilities $C$ as the following formula,

\begin{equation}\label{eq:z_share}
 \bm{z_{s}} = \frac{\bm{C}^T\bm{h}^3}{\bm{C}^T\mathbf{I}}
\end{equation}
where $\mathbf{I}$ is a $P\times 1$ all-ones vector, size of $\bm{z_{s}}$ is $K\times L$, 
the denominator is the normalization term. Finally, $\bm{z_{s}}$ is quantized into $\bm{z_{sQ}}\in [-1,1]^{K\times L} $ for storage. Later in the decoding procedure, the shared latent variables $\bm{z_{sQ}}$ for each cluster are merged by each image patch's soft label $\bm{C}$ to generate distinct latent variables $\bm{z_{f}}^{(3)}$ for each image patches as follows,
\begin{equation}
    \bm{z_{f}}^{(3)} = \bm{C}\bm{z_{sQ}}
\end{equation}
The latent variables $\bm{z_{f}}^{(3)}$ generated following the above process contains shared characteristics among patches for each cluster. As a result, our hierarchical latent variable framework is no longer independent between patches compared to the original L3C framework.


Finally, we define an auxiliary objective function that guides the learning process of $\bm{C}$ and $\bm{z_{sQ}}$. The aim of this objective function is to reconstruct the quantized latent variable $\bm{z_Q}^{(2)}$ in previous scale by $\bm{z_{f}}^{(3)}$. The loss $L_{cluster}$ is defined as follows,
\begin{equation}
    L_{cluster} = -\log_2 p(\bm{z_Q}^{(2)}|\bm{z_{f}}^{(3)})
\end{equation}
where $p$ is a parameterized discrete logitstic mixture model used in \cite{mentzer2019practical}.

\subsection{Loss Function}
For the hierarchical latent variable models in lossless image compression, the final compressed length consists of the length of compressed residual and latent variables will be modeled and minimized jointly. Therefore, the loss functions $L$ can be written as,

\begin{equation}\label{eq:loss}
    \begin{split}   L =
\underbrace{-\log_2 p(\bm{r}|{\bm{z_Q}^{(1)}},\bm{z_f}^{(1)})}_{L_{\bm{r}}} 
\underbrace{- \log_2 p({\bm{z_Q}^{(1)}}|{\bm{z_Q}^{(2)}},\bm{z_f}^{(2)})}_{L_{\bm{z_Q}^{(1)}}}
\underbrace{- \log_2 p({\bm{z_Q}^{(2)}}|\bm{z_f}^{(3)})}_{L_{cluster}} + L_{raw}  
\end{split}
\end{equation}
where all $p$ in Eq.\eqref{eq:loss} are parametrized by the discrete logitstic mixture model used in \cite{DBLP:journals/corr/SalimansKCK17} with 10 mixtures, $L_{\bm{r}}$ and $L_{\bm{z_Q}^{(1)}}$ are the length of compressed residual $\bm{r}$ and the quantized latent variable $\bm{z_Q}^{(1)}$, respectively. $L_{cluster}$ is the length of $L_{\bm{z_Q}^{(2)}}$. $\bm{C}$ and $\bm{z_{sQ}}$, $L_{raw}$ is the length of $\bm{C}$ and $\bm{z_{sQ}}$ which are stored uniformly.

\subsection{Compression}
Given the probability distribution modeled by our hierarchical latent variable framework, we encoded images using the Torchac \cite{mentzer2019practical} encoder, which is a kind of arithmetic entropy coder. We summarize the pipeline of GOLLIC to compress one image. (1) Encoding: The image after preprocessing will be passed through the encoder of our hierarchical latent variable model. The outputs are soft labels $\bm{C}$, shared latent variables $\bm{z_{sQ}}$ generated by the clustering module, and hierarchical latent variables $\bm{z_Q}^{(2)}$, $\bm{z_Q}^{(1)}$. The soft labels and shared latent variables are encoded according to the uniform distribution and then stored. The hierarchical latent variables $\bm{z_Q}^{(2)}$, $\bm{z_Q}^{(1)}$ and $\bm{r}$ are stored with length $L_{cluster}$, $L_{\bm{z_Q}^{(1)}}$, and $L_{\bm{r}}$ respectively. (2) Decoding: During decoding, $\bm{C}$ and $\bm{z_{sQ}}$ are decoded firstly from bitstream. Then they are used to reconstruct $\bm{z_{f}}^{(3)}$. Then $\bm{z_{f}}^{(3)}$ are fed into the decoder to obtain $\bm{z_{f}}^{(2)}$ and the conditional probability of $p(\bm{z_Q}^{(2)}|\bm{z_{f}}^{(3)})$. After entropy coder decode $\bm{z_Q}^{(2)}$ with the conditional probability, the reconstructed $\bm{z_{f}}^{(2)}$ and $\bm{z_Q}^{(2)}$ are fed into the decoder in next layer to to decode $\bm{z_Q}^{(1)}$ and $\bm{r}$ as shown in Fig.\ref{fig:gollic_framework}.

%% file: content/experiment.tex
\section{Experiment}
\subsection{Dataset}
We conduct experiments based on the following three high-resolution benchmark datasets, DIV2K. CLIC.pro and CLIC.mobile. 
DIVerse 2K resolution high quality images dataset (DIV2K) \cite{Agustsson_2017_CVPR_Workshops} is a  benchmarking dataset of single image super-resolution in NTIRE 2017 SR challenge and used in validating the performance of high-resolution image compression in recent works of literature of lossless compression. This dataset includes 900 2K resolution RGB images, divided into 800 training and 100 test data. In our experiment, we train our model based on 800 high-resolution images (640 for training, 160 for validation). We evaluate our model based on 100 test images.
CLIC.mobile and CLIC.professional (CLIC.pro) are provided by Workshop and Challenge on Learned Image Compression \cite{CLIC}. CLIC.mobile consists of 61 images, and CLIC.pro provide 41 images for validation. We evaluate CLIC.mobile and CLIC.pro's validation data using the pre-trained model trained in DIV2K. 
\subsection{Implementation Details}
In our experiment, the patch size is $128\times 128$.When cropping, the constant value 0 will be padded on the boundary of patch. We set channels of the latent variable $C_f = 64$, $C_d = 5$, the number of clusters  is $K = 5$. The details of architecture  is shown in supplementary material.  We trained our model 50 epochs with a batch size equal to 1 on training dataset. RMSProp optimizer was used to optimize this model with initial learning rate of $10^{-4}$ and decay it by the factor of 0.5 every 10 epochs. We implemented our model based on Pytorch and ran all the experiments by the machine equipped with an TESLA V100 with 32GBs of memory.

We evaluate our compression performance by bits per sub-pixel (bpsp). Each pixel contains three subpixels in RGB channels for a color image, and bpsp without compression is 8. For lossless compression, lower bpsp means a better compression ratio.  
\subsection{Compression Performance}
We evaluate compression performance of GOLLIC with engineered codecs, PNG \cite{Boutell1997PNGN}, JPEG2K \cite{rabbani2002book}, WebP \cite{WebP}, FLIF \cite{Sneyers2016FLIFFL}, BPG \cite{BPG} and deep learning models, L3C \cite{mentzer2019practical}, RC \cite{Mentzer2020LearningBL}, SReC \cite{cao2020lossless} based on three high-resolution image datasets. The result is shown in Table \ref{table:bpsp}. The experimental result shows that  GOLLIC outperforms the most engineered formats except FLIF. We achieve a higher compression ratio than two latent variable models, L3C and RC. Note that the normal result of L3C, SReC, and RC showed in Table \ref{table:bpsp} are trained on the OpenImage dataset, which contains 213487 images. However, the performance of our model trained on only 800 high-resolution images is still competitive with those models. When we train SReC and L3C on DIV2K (result denoted with $\star$,$\dagger$), our model outperforms their result significantly. This result indicates that modeling long dependencies on the high-resolution image is efficient.
\begin{table}[h]
\centering
\setlength{\tabcolsep}{0.35cm}{ 
\begin{tabular}{lcccc}
\toprule
\textit{bpsp} &DIV2K  &CLIC.pro &CLIC.mobile \\
\midrule
PNG \cite{Boutell1997PNGN} & $4.23$ &$3.99$ & $3.89$\\
JPEG2K \cite{rabbani2002book} & $3.12$   & $3.00$ &$2.72$\\
WebP \cite{WebP} &$3.17$ & $3.01$&$2.77$ \\
FLIF \cite{Sneyers2016FLIFFL}    &$\bm{2.91}$ & $\bm{2.78}$ &$\bm{2.49}$\\
BPG \cite{BPG} & $3.28$ & $3.08$&$2.84$ \\
\midrule
L3C \cite{mentzer2019practical} & $3.094$/$3.409^{\star}$ & $2.944$& $2.639$  \\
SReC \cite{cao2020lossless} &  $\bm{2.822}$ / $3.374^{\star}$/$3.338^{\dagger}$ & $\bm{2.660}$/$3.194^{\star}$/$3.160^{\dagger}$ &$\bm{2.440}$/${2.961}^{\star}$/$2.884^{\dagger}$ \\
RC \cite{Mentzer2020LearningBL} & 3.079 & 2.933&2.538 \\
\textbf{GOLLIC (Ours)}&$\bm{3.073}^{\star}$&$\bm{2.829}^{\star}$&$2.620^{\star}$ \\
\bottomrule
\end{tabular}}
\caption{Compression performance of the proposed GOLLIC based on three benchmarking high-resolution datasets compared to traditional engineered approaches and deep learning methods. The performance is measured in bpsp. $\star$ denotes the model trained on DIV2K, and the data has the same preprocessing with GOLLIC. $\dagger$ denotes the model trained on DIV2K without preprocessing. Normal results cited from L3C, SReC and RC are obtained by training on OpenImage. }
\label{table:bpsp}
\end{table}

\subsection{Inference Time}
We report the inference time of GOLLIC compared to SReC \cite{cao2020lossless} on GPU device Tesla V100. We measure the average inference time of models on DIV2K 100 test data. The inference time of GOLLIC is 1.64 second per image and that of SReC is 2.210 s. Thus, GOLLIC is faster than SReC, which has been much more lightweight than other latent variable models.

\subsection{Visualization of Clustering Result}
\begin{figure}[h]
    \centering
    \includegraphics[scale=0.7]{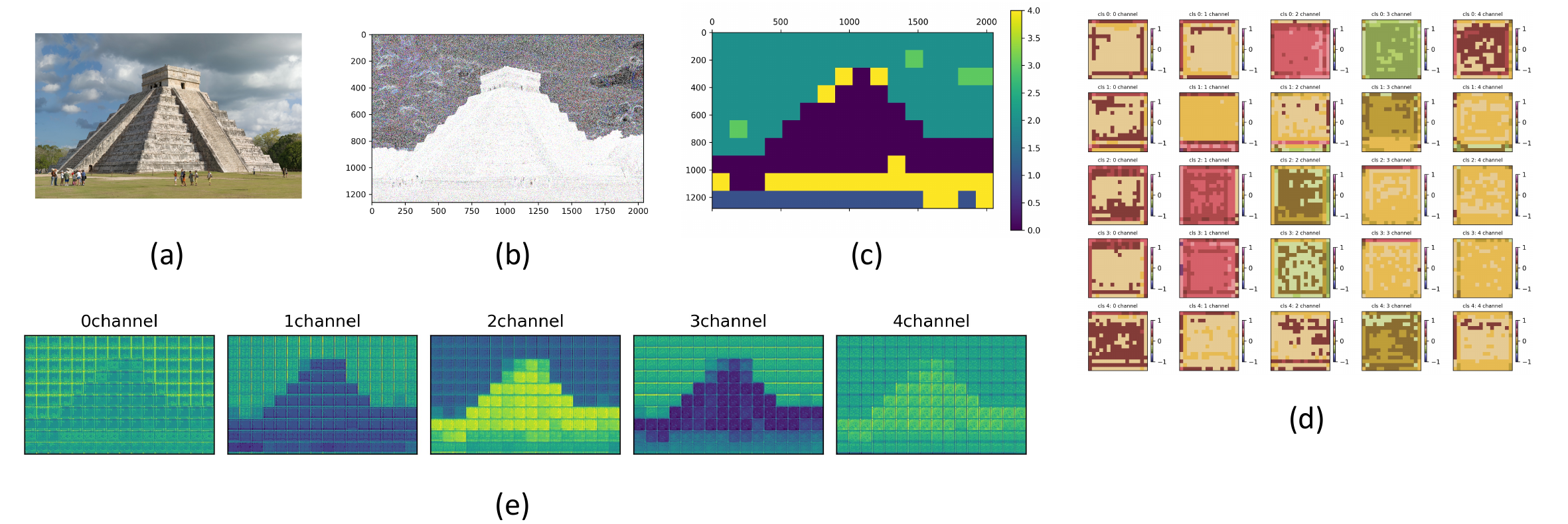}
    \caption{Visualization result of clustering and the shared latent variables. (a) Original Iamge. (b) The prediction residuals after preprocessing. (c) Clustering result. (d) the shared latent variables $\bm{z_{sQ}}$ on 5 clusters . (e) the reconstructed latent variable $\bm{z_f}^{(3)}$}
    \label{fig:vis_result}
\end{figure}

The visualization result of the clustering result and shared features are shown in Fig.\ref{fig:vis_result}. The clustering result visualizes the cluster on which each patch has the max probability. The result shows that our model has the power to cluster similar patches inside images, even on prediction residuals.

\subsection{Ablation Study}
\subsubsection{Patch size}
We investigate the effect of patch size on GOLLIC's performance on DIV2K dataset. We trained difference models based on three patch size, $128\times 128$, $64\times 64$ and $32\times 32$. Those models have trained on DIV2K dataset with 100 epochs. The results in Table \ref{table:patchsize} show that the patch size $128\times 128$ obtains the best performance. When we reduce patch size, the performance does not show many advantages. One possible explanation is that the smaller patch size of one image produces more patches, and the number of patches increases the difficulties of clustering.

\begin{table}[h]
\centering
\setlength{\tabcolsep}{18pt}{ 
\begin{tabular}{lc}
\toprule
 Patch size & bpsp  \\
\midrule
$128\times 128$& \textbf{3.073}\\
$64\times 64$ & 3.114\\
$32\times 32$& 3.206 \\
\bottomrule
\setlength{\belowbottomsep}{2cm}
\end{tabular}}
\caption{Performance of GOLLIC based on three scales of patch on DIV2K test data.}
\label{table:patchsize}
\end{table}

\subsubsection{Number of Clusters}
Cluster number $K$ is a key hyper-parameters to the compression performance. As $K$ approaches the number of patches, our model will fall back to the patch-independent models. When $K$ equals 1, the shared latent variable will collapse. Meanwhile, the enormous $K$ will cause more shared latent variables and cluster indexes to be stored. We explore the effect of several clusters by training models based on $1,5,10,20$ clusters on the DIV2K dataset with cropping patch size  $128\times 128$. The comparison results in Table \ref{table:cluster} shows that the model trained on only 1 cluster obtained the worst performance; 5-cluster model is the best. When the cluster number is larger than 5, the performance does not show much improvement. Combining the visualization result, we guess that the clustering module tent to group the critical characteristics inside the image, and 5 clusters are enough to describe them. We provide more ablation studies in Appendix \ref{app:ablation}.

\begin{table}[h]
\centering
\setlength{\tabcolsep}{0.15cm}{ 
\begin{tabular}{lccccc}
\toprule
Clusters & total bpsp & $L_{\bm{r}}$ & $L_{\bm{z_Q}^{(1)}}$&$L_{cluster}$& $L_{raw}$\\ 
\midrule
1& 3.644&3.431&0.169&0.430&0.001\\
5 & \textbf{3.073}&2.989&0.029&0.049&0.006\\
10& 3.325&3.228&0.033&0.050&0.013 \\
20& 3.122&3.032&0.006&0.050&0.026 \\
\bottomrule
\setlength{\belowbottomsep}{2cm}
\end{tabular}}
\caption{Performance of GOLLIC based on different number of clusters on DIV2K test data.}
\label{table:cluster}
\end{table}

%% file: content/conclusion.tex
\section{Conclusion}
This paper proposes a hierarchical latent variable model with a global context that captures the long-term dependencies inside the image, achieved by a self-supervised clustering module. Such long-term dependencies become important as the resolution of images increases. Due to limited memory or computational complexity, one must crop high-resolution images into fixed-size image patches for compression. Each image patch only represents a small part of semantic objects in the original picture. Therefore the long-term correlation between cropped patches shall not be neglected. Compared to other latent variable models assuming independency among cropped image patches, our model captures these long-term correlations as shared latent variables among image patches grouped by self-supervised clustering. Such shared latent variables benefit the data compression task by summarizing and removing the redundant and repeated patterns in each patch's latent variables. Experimental results also show that our model improves compression ratio compared to the engineered codecs and deep learning models on three benchmark high-resolution image datasets.


We remark that our proposed method may not necessarily have obvious advantages when applied to a low-resolution image. The reason is that long-term dependency between cropped patches is not as important as the dependency in high-resolution images. Contrarily, we expect our proposed method to show more obvious advantages over other latent variable models for images with higher resolution.

%% file: content/acknowledge.tex
\section*{Acknowledgement}
We would like to thank Candi Zheng for discussion and comments on the manuscript. We also thanks Zheting Dong for discussion on losselss compression. The work of Yuan Lan was supported by Huawei PhD fellowship.
The work of YX was partially supported by HKUST IEG19SC04, Minieye company through a research project at HKUST Shenzhen Institute, and the Project of Hetao ShenzhenHKUST Innovation Cooperation Zone HZQB-KCZYB-2020083.

%% file: content/appendix.tex
\appendix
\appendixpage

\section{Details of Model} \label{app:network}
\subsection{Preprocessing}
In main text, we mention that the original images are preprocessed by a reversible color transform \cite{gormish1997lossless}. We represent this integer transform which convert RGB to YCrCb  as follows,
\begin{align*}
        &Y = round\left\{\frac{R+2G+B}{4}\right\}, \\
    &Cr = R-G, \\
    &Cb = B-G. 
\end{align*}
The reverse transform which recover RGB from YCrCb is written as,
\begin{align*}
        &G = Y - round\left\{\frac{Cb+Cr}{4}\right\}, \\
    &R = Cr + G, \\
    &B = Cb + G. \\
\end{align*}
Besides this, when cropping the patches, we pad the boundary with constant 0 such that the image size is the integer multiple of patch size.
Further, as mentioned in the main text, when loading the data, the batch size is set to 1. In each step during training, we input the whole high-resolution image with size $(P,3, N, N)$ into the network, $N$ is patch size, and $P$ is the number of patches.

\subsection{Details of Architecture}
Now we describe the details of architecture in our models. GOLLIC contains three-layer encoders and decoders, we denote them as Encoder 1,Encoder 2,Encoder 3, Decoder 3,Decoder 2,Decoder 1. The self-supervised clustering module is embedded into Encoder 3. 
The modules inside Encoder $i$ ($i=1,2$) are: 
\begin{itemize}
    \item a Head $i$, which is a 2-dim convolution layer with input channel = 64 (for Head 1, input channel=3), output channel = 64, kernel size=3, padding =1, stride=1. 
    \item a downsampling layer, which is a 2-dim convolution layer with input channel = 64, output channel = 64, kernel size=5, padding =2, stride=2. 
    \item an eight-layer ResNet Block, which contains 8 ResNet blocks which map 64 input channels into 64 output channels. The activation function inside block is ReLU.
    \item a 2-dim convolutional layer with input channel = 64, output channel = 5, kernel size=3, padding =1, stride=1. 
    \item a quantizer used in \cite{mentzer2019practical}. We refer readers to \cite{mentzer2019practical} for the details. In our models, the quantization level is 25, hyper-parameter $\sigma_q = 2$, the data is quantized into $[-1,1]$.
\end{itemize}

As for Encoder 3 contains self-supervised clustering modules, its architectures are as follows,
\begin{itemize}
    \item a Head 3 with input channel = 64, output channel = 64, kernel size=3, padding =1, stride=1. 
    \item a Classification Head contains 
    \begin{itemize}
        \item a 2-dim convolution with input channel = 64, output channel = 64, kernel size = 3, padding =1, stride=1, a ReLU function, a 2-dim convolution with input channel = 64, output channel = 5, kernel size = 5, padding =2, stride=2.
        \item a fully-connected layer with input size = 320, output size = 5.
        \item a softmax layer.
    \end{itemize}
    \item an eight-layer ResNet Block, which contains 8 ResNet blocks which map 64 input channels into 64 output channels. The activation function inside block is ReLU.
    \item a quantizer as metioned above. In our models, the quantization level is 25, hyper-parameter $\sigma_q = 2$, the data is quantized into $[-1,1]$.
\end{itemize}

Next, we describe the modules inside the Decoder $i, i=1,2,3$ as follows,
\begin{itemize}
    \item a Head $i$, which contains two convolution layers. Layer 1 is a 2-dim convolution with  input channel = 5, output channel = 64, kernel size = 1, padding =0, stride=1. Layer 2 is a 2-dim convolution with  input channel = 64, output channel = 64, kernel size = 1, padding =0, stride=1. 
    \item  an eight-layer ResNet Block, which contains 8 ResNet blocks which map 64 input channels into 64 output channels. The activation function inside block is ReLU.
    \item an upsampling layer, which contains a convolution layer with input channel = 64, output channel = 256, kernel size = 3, padding =1, stride=1, a PixelShuffle \cite{DBLP:journals/corr/ShiCHTABRW16} with upscale factor = 2, and an identity function. 
\end{itemize}
In our model, we use the discritized logistic mixture model with 10 components proposed in \cite{DBLP:journals/corr/SalimansKCK17}. For the estimation of $p({\bm{z_Q}^{(1)}}|{\bm{z_Q}^{(2)}},\bm{z_f}^{(2)})$ and $p({\bm{z_Q}^{(2)}}|\bm{z_f}^{(3)})$, we did note share the parameters among channels and the number of parameters for each component inside each channel is 3. And for $ p(\bm{r}|{\bm{z_Q}^{(1)}},\bm{z_f}^{(1)})$, we share the parameters among channels and the  the number of parameters for each component inside each channel is 4.

\section{More Ablation Studies}\label{app:ablation}
We add more ablation studies based on three effects, patch size, number of clusters, and scale of models. For effect 1 of patch size, we set patch size from $16\times 16$ to $256\times 256$. As shown in Table \ref{table:ablation}, the performance improves with increasing patch size from $16$ to $128$. However, the patch size is large enough ($256$), and the performance does not improve further. During the experiments, the whole image is cropped into fewer patches when the patch size is larger. Too few patches may not provide useful information for clustering. Thus, the performance with patch size 256 is worse than that of patch size 128. Next, we analyze the result of effect 2, cluster number. We found that more cluster numbers do not improve performance (such as 50 clusters).
Meanwhile, only 1 cluster will cause the model to lose its expressiveness. Models trained with a 5-cluster achieve the best performance. Finally, we test the effect of the model scales on the compression results. We use model scales to denote the number of encoder-decoder in hierarchical models. We note that as the scales of the model increase, the performance is better. Note that when scale=1, the auxiliary objective function in our clustering module does not work since $\bm{z_Q}$ in the previous layer does not exist. Thus, the clustering module collapsed in the model with a scale =1.
\begin{table}[H]
\setlength{\tabcolsep}{0.25cm}{ 
\begin{tabular}{lcccccccc}
\toprule
&Patch size& Clusters &Scale & bpsp& $L_{\bm{r}}$ & $L_{\bm{z_Q}^{(1)}}$&$L_{cluster}$& $L_{raw}$ \\
\midrule
\multirow{1}{5em}{Baseline} &-   & -   &-&$3.475$&3.475&-&-&-\\
\midrule
\multirow{5}{5em}{Effect 1: patch size} &$\mathbf{16 \times 16}$   & 5   &3&3.486&3.234&0.055&0.029&0.168\\
&$\mathbf{32 \times 32}$  & 5   &3&3.206&3.049&0.040&0.075&0.043\\
&$\mathbf{64 \times 64}$  & 5& 3&3.114&3.023&0.025&0.054&0.012\\
&$\mathbf{128 \times 128}$  & 5&3&\textbf{3.073}&2.989&0.029&0.049&0.006\\
&$\mathbf{256 \times 256}$  & 5&3&3.087&3.003&0.014&0.056&0.012 \\
\midrule
\multirow{5}{6em}{Effect 2: cluster number} &$128 \times 128$   & \textbf{1}   &3&3.644&3.431&0.169&0.430&0.001\\
&$128 \times 128$  & \textbf{5}   &3&\textbf{3.073}&2.989&0.029&0.049&0.006\\
&$128 \times 128$  & \textbf{10}& 3&3.325&3.228&0.033&0.050&0.013\\
&$128 \times 128$  & \textbf{20}&3&3.122&3.032&0.006&0.050&0.026\\
&$128 \times 128$  & \textbf{50}&3&3.291&2.787&0.338&0.100&0.066 \\
\midrule
\multirow{3}{6em}{Effect 3: model sclaes} &$128 \times 128$   & 5   &\textbf{1}&3.616&3.559&-&-&0.057\\
&$128 \times 128$  & 5   &\textbf{2}&3.490&3.269&0.204&-&0.017\\
&$128 \times 128$  & 5& \textbf{3}&\textbf{3.073}&2.989&0.029&0.049&0.006\\
\bottomrule
\end{tabular}}
\caption{Ablation studies on three effects, patch size, number of clusters, and scale of models trained on DIV2K training data, evaluated on DIV2K test data. The baseline result is calculated by first-order entropy $H$ of data after preprocessing, where $H = -\sum_{i=1}^{n}p(x_i)\log_2 p(x_i)$, $x_i$ is the symbol inside finite alphabet $\mathcal{A}$, $p(x_i)$ is probability of symbols $x_i$ occurring.}
\label{table:ablation}
\end{table}
\section{Comparison to Benchmarking Approaches} \label{app:benchmark_exp}
The codes of L3C \cite{mentzer2019practical} and SReC \cite{cao2020lossless} are from \url{https://github.com/fab-jul/L3C-PyTorch} and \url{https://github.com/caoscott/SReC}, respectively. In our experiments, the models of L3C and SReC are the default setting, as shown in the codes in the above links. We train these two models with 50 epochs in the same machine with GOLLIC, and their training curves converge after 50 epochs.